# Anisotropic Schottky-barrier-height in high-symmetry 2D WSe$_2$: Momentum-space anisotropy

Nuo Xu, Xiao-Lin Zhao, Meng-Xue Ren, Ke-Xin Hou, Xiao-huan Lv, Rui-Ning Wang, Xing-Qiang Shi*, and Jiang-Long Wang*

*Key Laboratory of Optic-Electronic Information and Materials of Hebei Province, Hebei Research Center of the Basic Discipline for Computational Physics, College of Physics Science and Technology, Hebei University, Baoding 071002, P. R. China*

*E-mails: shixq20hbu@hbu.edu.cn, jlwang@hbu.edu.cn

Abstract: It is usually supposed that only low-symmetry two-dimensional (2D) materials exhibit anisotropy, here we show that high-symmetry 2D semiconductors can show significant anisotropy in momentum space due to the band structure anisotropy in *k*-space. The basic reason is that different *k*-points in the Brillouin zone have different symmetry. Using 2D semiconductor WSe$_2$ as the example, we construct lateral heterostructures with zigzag and armchair connections to 2D metal NbSe$_2$, and the electronic structure and contact characteristics of these two connections are analyzed. It is found that both connections exhibit *p*-type Schottky barrier height (SBH) but the sizes of SBH are very different (of 0.03 eV and 0.50 eV), mainly because the band-edge energies of WSe$_2$ are different along the two mutually perpendicular directions in momentum space. There are two factors contributing to the SBH anisotropy: one is the different interface structure and the other is the band edge anisotropy of the 2D semiconductor WSe$_2$. Since the two interface structures give only a difference in interface potential change by less than 0.1 eV, the SBH variation of ~0.47 eV is mainly from the band structure anisotropy in momentum-space. So, high-symmetry 2D materials may exhibit highly anisotropic electronic states in momentum space and this affects the transport properties. Our current work extends the research field of 2D material anisotropy to 2D materials with high real-space symmetry, thus greatly expands the candidate materials for anisotropic studies and provides new guidance for optimizing the performance of 2D material devices via controlling transport directions.

# I. INTRODUCTION

Two-dimensional (2D) materials have attracted great attention due to their excellent physical properties [1-3] and (photo)electric applications [4-6], and are regarded as an ideal candidate system to continue Moore's Law in the post-silicon-based semiconductor era [7-11]. The in-plane anisotropy of 2D materials is mainly manifested in the angular dependence of optical, electrical and transport properties. Due to their potential application in the next-generation anisotropic multifunctional devices, low-symmetry 2D materials with intrinsic anisotropy (in real-space crystal structures) have attracted extensive research interest, including their intrinsic properties, transport property regulation and device optimization [12-17]. However, it is usually supposed that only low-symmetry 2D materials exhibit anisotropy; and there is a lack of study on whether high-symmetry 2D semiconductors can show significant anisotropy in momentum space.

In-plane anisotropic 2D materials with low-symmetry crystal structures have gained extensive study. Typical examples includes orthorhombic black phosphorus [13], monoclinic T'-WTe$_2$ [18], triclinic ReS$_2$ [19], and the corrugated pentagonal PdPSe [14]. These studies have demonstrated the uniqueness of anisotropy in material properties and device applications, facilitating multifunctional manipulation and experimental measurements. For example, Miao *et al.* studied the in-plane anisotropy of ReS$_2$ and proposed a method for fabricating highly efficient logic inverters, offering a novel perspective for the application of 2D materials in integrated circuits by utilizing lattice orientation as a new device design parameter [12]. The above anisotropy of 2D materials refers to in-plane anisotropy, and the research works on anisotropy are based on real-space geometric structure anisotropy. Such as theoretical studies have revealed the anisotropic band structure of ReS$_2$ (layer group $p\bar{1}$): namely, as a result of real-space anisotropy, the irreducible Brillouin zone (BZ) contain half of BZ instead of $\frac{1}{12}$ BZ as usual for the high-symmetry MoS$_2$ or WSe$_2$ (layer group $p\bar{6}m2$), and three times of ***k***-paths are needed to represent the full band structure with anisotropy in ***k***-space [17]. For ReS$_2$, the ***k***-space anisotropy is a result of real-space anisotropy, and it is usually supposed that only low-symmetry 2D materials exhibit anisotropy. Here in the current work, we show that high-symmetry 2D semiconductors (such as WSe$_2$) can show significant anisotropy in ***k***-space due to the band-edges anisotropy in momentum space. The basic reason is that different ***k***-points in the BZ have different symmetry, which may result in different orbital hybridization, different orbital compositions, and different band edge energy-positions along different directions in momentum space. This is discussed in detail in Note SI of Supplemental Material [20], which includes Refs. [21, 22].

We constructed lateral heterojunctions of metallic NbSe$_2$ and semiconducting WSe$_2$ connected along zigzag and armchair edges. By calculating the electronic structures and contact properties of these two types of heterojunctions, we found that the *p*-type Schottky-barrier-height (SBH) for both junctions, but the SBH for the zigzag junction is small (0.03 eV) while that for the armchair one is large (0.50 eV). This difference arises from the distinct band dispersion characteristics along the two directions, demonstrating that high-symmetry 2D materials can exhibit strongly anisotropic electronic states in momentum space. There are two factors contribute to the SBH anisotropy: one is the different interface structure between metallic NbSe$_2$ and semiconducting WSe$_2$ and the other is the band edge anisotropy of the 2D semiconductor WSe$_2$. Since two interface structures give only a difference in potential changes by less than 0.1 eV (details see below), the SBH variation of ~0.47 eV is mainly from the band structure anisotropy in momentum-space. Furthermore, we calculated the transmission coefficients of these two junctions at zero bias and confirmed the SBH difference between the two transport directions. Our current work 1) demonstrates the band structure anisotropy in momentum-space for high-symmetry 2D materials, probes the different origins of band structures anisotropy and demonstrates its application for SBH tuning; 2) extends the research field of 2D material anisotropy to 2D materials with high real-space symmetry, thus greatly expands the candidate materials for anisotropic studies; and 3) provides new insights for 2D material devices optimization, emphasizing the importance of the electrode connection direction.

## II. COMPUTATIONAL METHODS

First-principles density-functional theory (DFT) calculations [23] were performed using the Vienna *Ab initio* Simulation Package (VASP) [24-26]. The projector augmented-wave (PAW) [27, 28] potential was adopted to describe the core electrons. The exchange-correlation functional adopted the generalized gradient approximation (GGA) [29] in the Perdew-Burke-Ernzerhof form (GGA-PBE). The valence electrons were described by plane-wave basis with an energy cut-off of 400 eV. The convergence criteria were $10^{-5}$ eV for the self-consistent energy calculation and 0.02 eV/Å for the Hellmann-Feynman force in the geometric optimization. For the Brillouin zone sampling, 8×4×1 Monkhorst-Pack scheme [30] *k*-points was used for the 1×√3 supercell WSe$_2$, 8×1×1 *k*-mesh was used for the zigzag NbSe$_2$/WSe$_2$ supercell and 1×4×1 $k$-mesh was used for the Armchair NbSe$_2$/WSe$_2$ supercell while 12×12×1 $k$-mesh was used for the 1×1 unit cell WSe$_2$. They all corresponding to the similar $k$-point density. The vacuum separation in the vertical direction was 15 Å to ensure that the interaction between periodic images is negligible. The transmission coefficients were calculated based on the non-equilibrium Green's function(NEGF) method as implemented in the QuantumATK software package [31]. A real-space grid density which was equivalent to a plane-wave kinetic energy cut-off of 105 Ha was used. The $k$-point grid mesh of 4×1×150 was adopted to sample the Brillouin zone for the electrodes of zigzag and Armchair NbSe$_2$/WSe$_2$ supercells. More on calculation details can be found in the related sections. In addition, the VASPKIT [32] and VESTA [33] software packages were used for data analysis and presentation.

## III. RESULTS

In the following, we take the monolayer H-phase WSe$_2$ with high-symmetry as an example to illustrate the concept of momentum-space anisotropy. Figure 1(a) is a top-view of WSe$_2$ with the 1×1 normal cell and the 1×√3 rectangular cell indicated. For the 1×1 cell, the two directions along the two basis vectors ($a_1$ and $a_2$) are equivalent. In contrast, for the 1×√3 cell, the directions of the two basis vectors are inequivalent ($x$ and $y$ with 90° between them). Namely, they are along the zigzag and armchair directions, respectively. Figure 1(b) is the corresponding Brillouin zones (BZs) of the two cells, where blue (red) represents the BZ of 1×1 (1×√3) cell, and the Γ-X (Γ-Y) direction of the rectangular BZ of 1×√3 cell is along the same direction to the Γ-K (Γ-M) direction of the hexagonal BZ of 1×1 cell. For the 1×1 cell, since the M-K′ path is parallel to the Γ-K path, to determine the band edge along band the Γ-K direction, the bands along both M-K and Γ-K paths need to be considered; similarly, for the 1×√3 cell, to determine the band edge along Γ-X direction, the bands along both Γ-X and Y-S paths need to be considered, since they are parallel, and for the Γ-Y direction, the bands along both Γ-Y and X-S paths need to be considered. It should be noted that for electron transport across the material, it can be along either the Γ-X or Γ-Y directions (or equivalently, along the Γ-K or Γ-M directions), and these two directions are anisotropic.

Figures 1(c, d) shows the band structure and its projection to the $d$-orbitals of W atom of WSe$_2$ with the 1×1 normal cell. It can be observed that the band edges are anisotropic along the Γ-K and Γ-M directions, which show two main differences: 1) the difference in band edge energy-positions [Fig.

1(c)] and 2) the difference in the orbital compositions [Fig. 1(d)]. Namely, the band edges are different both in energy and in orbital composition along the two directions. Specifically, the valence band (VB) edge along the Γ-K direction is higher in energy than the VB edge along the Γ-M direction as shown in Fig. 1(c). Since the band edges are mainly contributed by the $d_{z^2}$, $d_{xy}$ and $d_{x^2-y^2}$ orbitals of the transition metal atom [34], Fig. 1(d) shows the band structure projected to the $d$-orbitals of W atom. Figure 1(d) shows that the VB edge along the Γ-M (Γ-K) direction is mainly contributed by the out-of-plane $d_{z^2}$-orbital (in-plane $d_{xy}$ and $d_{x^2-y^2}$ orbitals). More projected bands with projections to the $s$, $p$, and $d$-orbitals are given in Fig. S2 of Supplemental Material [20], and Fig. 1(d) is derived from Figs. S1(a, b) since these orbitals contribute most to the band edge. So, the two directions in ***k***-space is anisotropic in both band edge energy-positions and in orbital compositions. Figures 1(e, f) are the same to Figs. 1(c, d) but with the 1×√3 rectangular cell: the two directions (Γ-K and Γ-M) now changes to Γ-Y and Γ-X directions [refer to Fig. 1(b) for the directions in BZs]; and the band edges at the K point of 1×1 cell in Fig. 1(c) is folded to the Λ point of 1×√3 cell in Fig. 1(e).

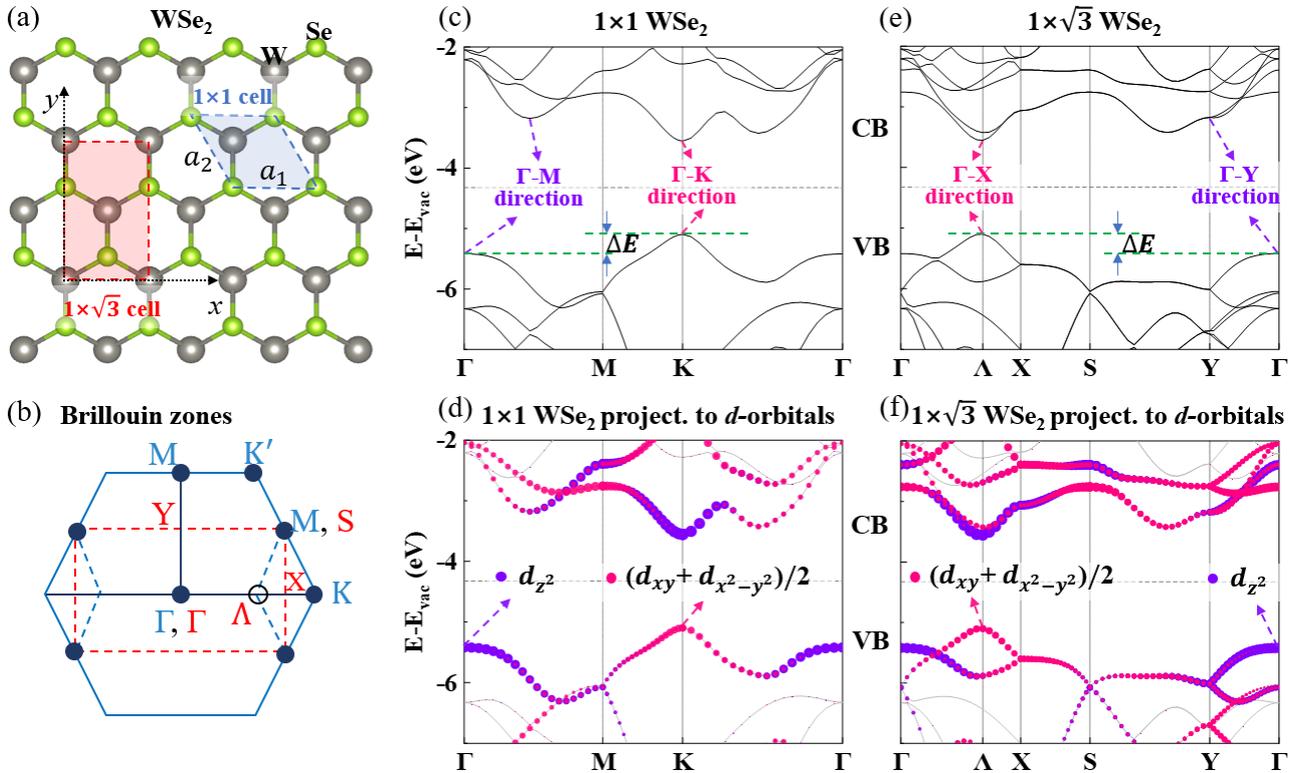

**FIG. 1. Anisotropic band structure and orbital composition in *k*-space for WSe$_2$: analyzed from real-space cells (1×1 normal cell and 1×√3 rectangular cell) and the corresponding Brillouin zones, band structures and band structures projected to *d*-orbitals.** (a, b) The 1×1 and 1×√3 cells and the corresponding Brillouin zones (BZ); blue and red represent 1×1 and 1×√3 cells, respectively; in (b), the Γ-X (Γ-Y) direction of the rectangular BZ of the 1×√3 cell is along the same direction to the Γ-K (Γ-M) direction of the hexagonal BZ of the 1×1 cell. (c, d) the band structures of WSe$_2$ with 1×1 cell: the valence and conduction band edges (VB edge and CB edge) along the Γ-M and Γ-K directions are labeled out with an energy difference of ΔE in (c); in (d), the band structure is projected to the in-plane and out-of-plane *d*-orbitals of W atom. (e, f) the same to (c, d) but with the 1×√3 cell: the band edges along the Γ-Y and Γ-X directions are labeled out with an energy difference of ΔE in (e); the band edges at the K point of 1×1 cell in (c) is folded to the Λ point of 1×√3 cell in (e) [refer to (b) for K and Λ points in BZs]. $E_{\text{vac}}$ in (c-f) denotes vacuum energy level and the VB and CB are labeled out.

In the following, the rectangular 1×√3 cell is used for the convenience of discussion of the anisotropic Schottky barrier height (SBH).

The metal NbSe$_2$ with a work function of 5.54 eV has been selected to form *p*-type lateral metal-semiconductor junctions (MSJ) with WSe$_2$ in the armchair and zigzag directions. The NbSe$_2$/WSe$_2$ junctions have been prepared in experiment [35, 36]. Figure 2 illustrates the lateral MSJ formed by WSe$_2$ and NbSe$_2$. The lattice constants of the two materials are similar and at the junction interface the lattice size is fixed to that of WSe$_2$. Note that this detail on how to handle lattice mismatch does not change the main conclusion of the current work. Two types of interfaces, armchair and zigzag interfaces, are considered. The literature [37] indicated that when the width of the semiconductor in the zigzag heterojunction is greater than 19 Å and the width of the semiconductor in the armchair heterojunction is greater than 22 Å, the SBH will be roughly unchanged. The structures we adopted are both greater than the above widths as indicated in Ref. [37].

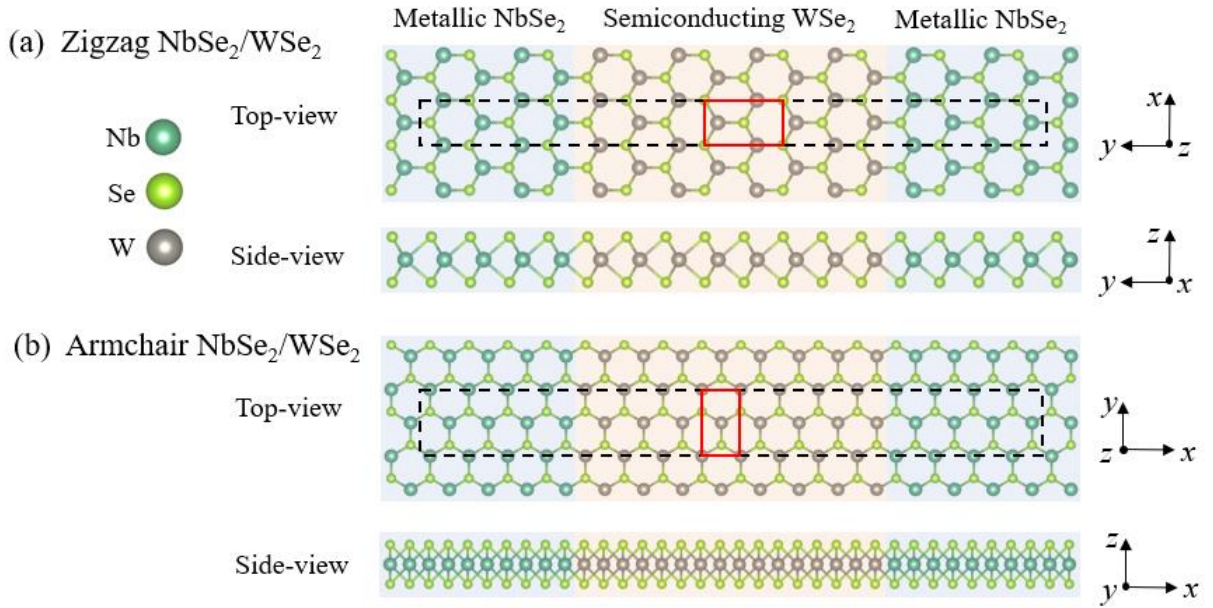

**FIG. 2. Side and top views of the lateral metal-semiconductor junctions (MSJ) of NbSe$_2$/WSe$_2$: (a) zigzag and (b) armchair junctions.** The red rectangular cell in top-views shows that the 1×√3 cell is along different orientations in the two junctions; the black supercell is the simulation cell for MSJ in DFT calculations.

The NbSe$_2$/WSe$_2$ lateral MSJ results in Schottky contacts with a finite SBH. Firstly, we discuss which of the two directions (parallel or perpendicular to the NbSe$_2$/WSe$_2$ interface) determine the anisotropic SBH for electron transport across the junction. For WSe$_2$ in lateral MSJ, the direction parallel to the NbSe$_2$/WSe$_2$ interface is periodic while for the direction perpendicular to the interface, the periodicity is disrupted by the interface potential. From the electron transport point of view, for metal-semiconductor junctions of 3D bulk materials, the transmission coefficient (T) across the metal-semiconductor interface is the function of incident electrons with energy E, lateral wave-vector $k_\parallel$, and $k_\perp \to k'_\perp$ (the normal wave-vector changes from $k_\perp$ to $k'_\perp$): namely, T(E, $k_\parallel$, $k_\perp \to k'_\perp$), which means only $k_\parallel$ are good quantum numbers and the $k_\perp$ can be scattered (or $k_\perp$ is not conserved), see for example literature [38]. For 2D materials, the "interface" becomes a "interline", and the $k_\parallel$ along lateral directions becomes the direction of the "interline". Above is the theoretical basis for the SBH anisotropy analysis in momentum-space. So, the SBH of WSe$_2$ in MSJ is determined by the band edge along the direction parallel to the interface. Namely, for the zigzag MSJ in Fig. 2(a), the band dispersion along the Γ-X direction should be considered; while for the armchair MSJ in Fig. 2(b), the band dispersion along the Γ-Y direction should be considered.

Our definition of SBH is based on the definition but with an extension of it to include the in-plane anisotropy. And the anisotropic SBH has been reported for low-symmetry 2D semiconductors in very recent literatures [39, 40]. The discussion for the extension of the traditional SBH definition to including the SBH anisotropy along different directions are given in Note SII of Supplemental Material [20]. Therefore, Figs. 3(a, b) display the band dispersion and orbital projection along the direction parallel to the zigzag and armchair interfaces for the SBH of the zigzag and armchair MSJs, respectively. Recall that Figs. 1(e, f) above show that the VB edge of WSe$_2$ is located at the Λ point along the Γ-X direction, and the orbital components consists of the in-plane $d_{xy}$ and $d_{x^2-y^2}$ orbitals of the W atom; and, along the Γ-Y direction the VB edge is at the Γ point with orbital components of out-of-plane $d_{z^2}$ orbital.

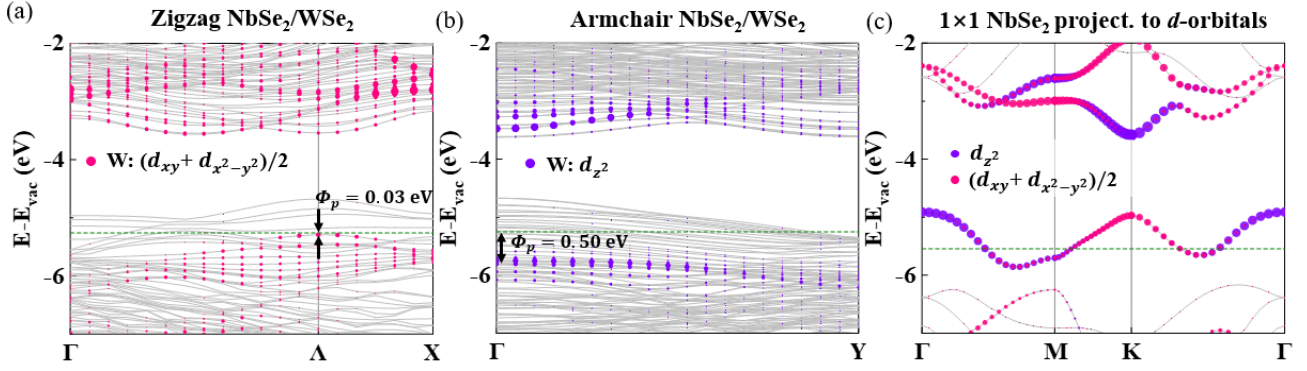

**FIG. 3. Anisotropic *p*-type SBH ($\Phi_p$) for the zigzag and armchair NbSe$_2$/WSe$_2$ MSJs.** (a, b) SBH determined from the projected bands along the periodic direction parallel to the interface of MSJ (see text). (c) The band structure of metallic NbSe$_2$ with 1×1 cell is projected to the in-plane and out-of-plane *d*-orbitals of Nb atom, whose Fermi level is close to the valence band of WSe$_2$. Dotted green line represents Fermi level.

Figures 3(a, b) show the projected band structure of the *d*-orbital of W atom in the zigzag and armchair NbSe$_2$/WSe$_2$ MSJs; Fig. 3(c) shows the band structure of NbSe$_2$ with 1×1 cell projecting to the in-plane and out-of-plane *d*-orbitals of Nb atom, whose Fermi level is close to the valence band of WSe$_2$. The anisotropic *p*-type SBH can be obtained using the following formulae:

$$\Phi_p(\text{zigzag}) = E_F - E_{\text{VB}}(\text{zigzag});$$
$$\Phi_p(\text{armchair}) = E_F - E_{\text{VB}}(\text{armchair}).$$

Here, $E_{\text{VB}}(\text{zigzag})$ and $E_{\text{VB}}(\text{armchair})$ represent the VB edge of WSe$_2$ in MSJ along the two directions; $E_F$ is the Fermi level of the metal NbSe$_2$ in MSJ. Figures 3(a, b) show that the $\Phi_p$ for the zigzag (armchair) MSJ is 0.03 eV (0.50 eV), demonstrating an significant anisotropic SBH. However, the junctions composed of zigzag and armchair interfaces have different interface structures, therefore, we calculate the interface potential change ($\Delta V$) of the two interfaces to analyze its effect on the anisotropic SBH, using the same method as that reported in literature [41]. The details are provided in Note SIII of Supplemental Material [20], which shows that the SBH anisotropy from the factor of different interface structure is less than 0.1 eV. As a result, the SBH variation of ~0.47 eV is mainly from the band structure anisotropy in momentum-space, the interface structures is not the dominate factor.

In order to confirm the significant anisotropic SBH proposed above, we use the QuantumATK package [31] to calculate the zero-bias transmission coefficients of zigzag and armchair NbSe$_2$/WSe$_2$/NbSe$_2$ MSJs. Figures 4(a, b) are the two-terminal device structures, in which the source and drain is the same metal electrode NbSe$_2$, and the central scattering region contains WSe$_2$ and a buffer electrode layer. The length of the scattering region is similar for both MSJs: 46.03 Å and 46.51 Å [refer to Figs. 4(a, b)]. Figures 4(c, d) compare the zero-bias transmission coefficients of the zigzag and armchair MSJs, from which the anisotropic SBH is confirmed. However, there is an energy gap underneath the small peak [as labeled out in Fig. 4(c)]: this is due to the 2D electrode of $NbSe_2$ is not a good metal and has an energy gap in this energy interval, as shown in Fig. S6 of Supplemental Material [20]. The band structure of $NbSe_2$ in Fig. S6(a) shows that there is an energy gap between a partially occupied band and a fully occupied band at about -6 eV relative to vacuum level (or below the Fermi level of $NbSe_2$). And we increased the channel length of $WSe_2$ by two times in zigzag and armchair $NbSe_2/WSe_2/NbSe_2$ MSJs to verify the small peak below the Fermi level in Fig. 4(c) is not an impurity state. The calculated transmission coefficient as shown in Fig. S7 of Supplemental Material [20]. The calculation result is similar to that in Fig. 4(c), which shows the small peak still exists, indicating that it is not an impurity state.

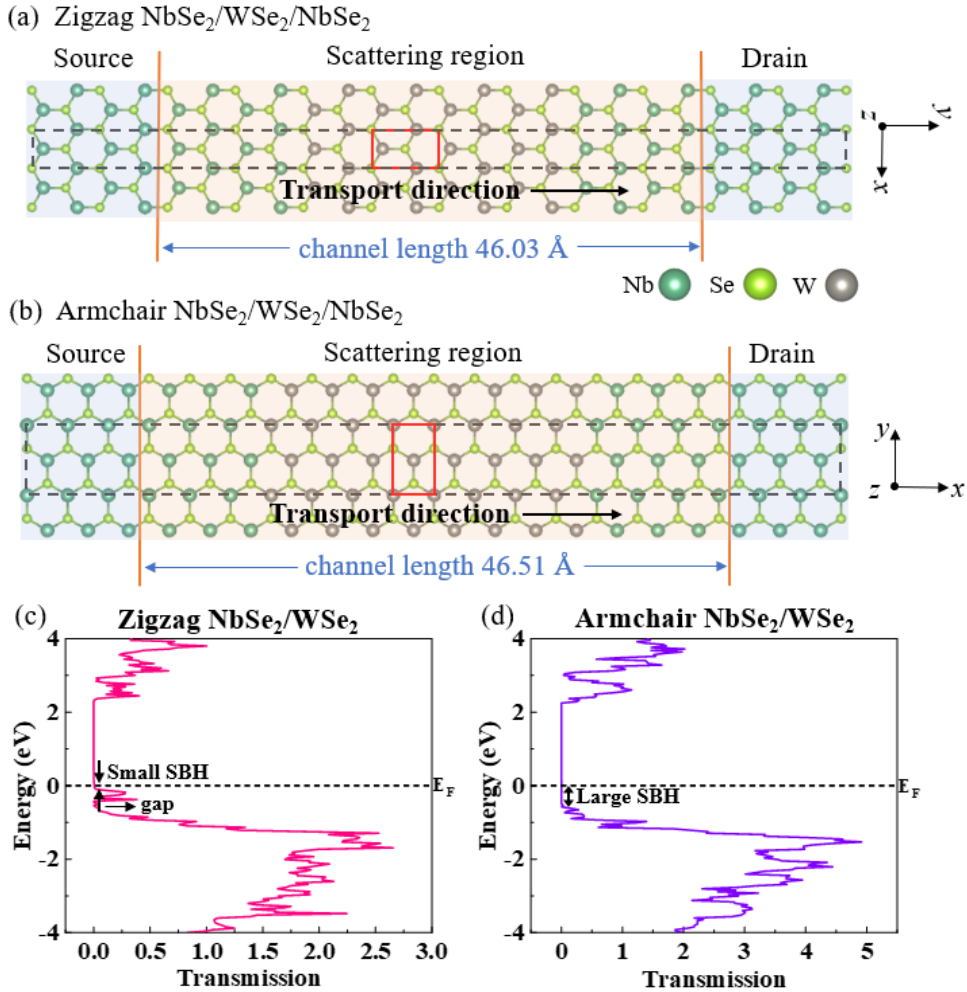

FIG. 4. **Anisotropic SBH confirmed from zero-bias transmission coefficients.** (a, b) Two-terminal device

structures of zigzag and armchair NbSe$_2$/WSe$_2$/NbSe$_2$ MSJs. (c, d) Zero-bias transmission coefficients of the two MSJs.

In the above, we estimate the SBHs from the *d*-orbitals of W and verify them with the transmission calculations. Table I lists the SBH values along different directions from *d*-orbitals of W atom and from transmission calculations. Table I shows that, the difference between SBHs along the Γ-M and Γ-K directions (ΔSBH) obtained from the two methods are similar.

TABLE I. The SBH values from *d*-orbitals of W atom and from transmission calculations. ΔSBH is the difference between SBHs along the Γ-M and Γ-K directions.

| Methods | SBH ($E_{Γ-K}$) | SBH ($E_{Γ-M}$) | ΔSBH ($E_{Γ-M}-E_{Γ-K}$) |
|---|---|---|---|
| *d*-orbitals of W atom | 0.03 | 0.50 | 0.47 |
| Transmission calculations | 0.1 | 0.58 | 0.48 |

## IV. DISCUSSION

It is usually supposed that only 2D materials with low symmetry exhibit significant anisotropy. The above results show that significant anisotropy in ***k***-space may also exhibit in 2D materials with high symmetry, including anisotropic VB edge energy, orbital character, and SBH, as summarized/sketched in Fig. 5(a). Figure. 5(a) sketches the SBH anisotropy for *p*-type and *n*-type contacts to WSe$_2$ assuming the metal contacts with suitable Fermi levels. To determine the band edge along Γ-X direction, the bands along both Γ-X and Y-S paths need to be considered, since the two paths are parallel [see Fig. 5(b)]. Similarly, for the Γ-Y direction, the bands along both Γ-Y and X-S paths need to be considered. A sketch of anisotropic electron transport for high-symmetry 2D materials with different contacts are given in Fig. S8 of Supplemental Material [20]. Figure S8(a) indicates the anisotropic SBH between Γ-X and Γ-Y directions, and Figs. S8(b, c) indicate the importance of direction of electrode connection on electron transport: one direction favor electron transport while the other direction does not. Finally, although our study focuses on edge contact (lateral MSJ), our findings of anisotropic SBH and hence anisotropic electron transport should not be limited to edge contact, but should also applicable for top contact and hybrid contacts with different directions of contact, as sketched in Figs. S8(d, e).

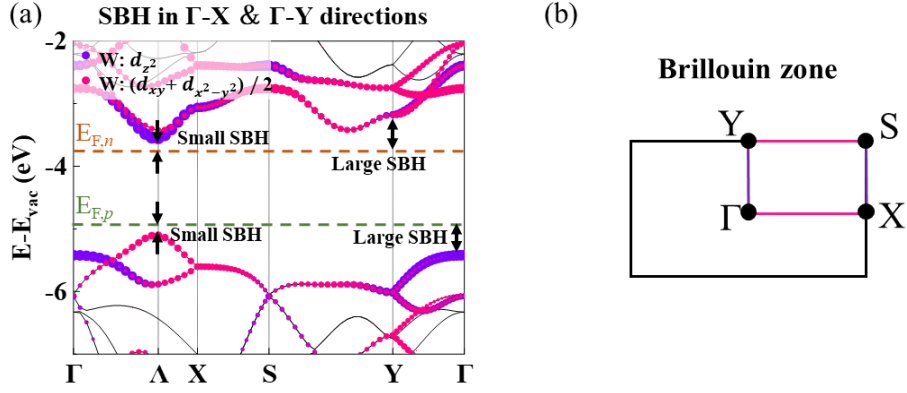

FIG. 5. (a) Anisotropic SBH of WSe$_2$ with *n*-type and *p*-type contact electrodes with suitable Fermi level, as sketched by E$_{F,n}$ and E$_{F,p}$, respectively. $E_{\text{vac}}$ denotes vacuum energy level. (b) The Brillouin zone used in (a). To determine the band edge along Γ-X direction, the bands along both Γ-X and Y-S paths need to be considered; similarly, for the Γ-Y direction, the bands along both Γ-Y and X-S paths need to be considered.

It should be note that the moment-space anisotropy is not limited to WSe$_2$. Figure 6 summarizes the anisotropic band-edge energies along Γ-M and Γ-K directions with 1×1 cell for other group VIB transition metal dichalcogenides (TMDs) monolayers in H-phase. Figure 6 shows that, for all of the group VIB TMDs, the energy values of VB and CB edges are anisotropic between Γ-M and Γ-K directions. The labeled numbers show the sizes of energy difference between the two directions: most of the energy differences are significant (from 0.3 to 0.5 eV), indicating that these high-symmetry 2D materials have strong electronic state anisotropy in momentum space for CB and/or VB.

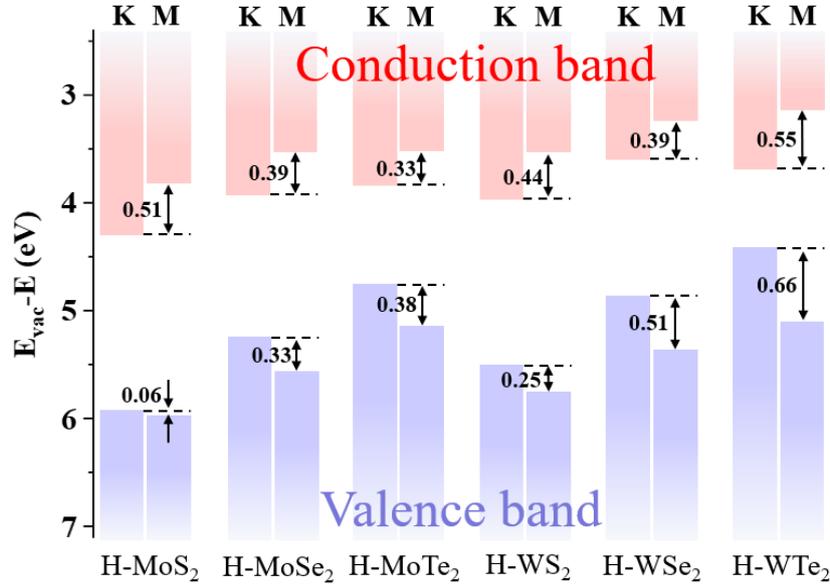

**FIG. 6. Summary of anisotropic band-edge energies along Γ-K and Γ-M directions with 1×1 cell for group VIB transition metal dichalcogenides (TMDs) monolayers in H-phase.** The "K" and "M" labeled on the top denote Γ-K and Γ-M directions, respectively; the band-edge values are calculated with PBE+SOC.

## V. CONCLUSION

We find that high-symmetry 2D semiconductors can show anisotropy in momentum space due to the different *k*-points in the BZ have different symmetry, which may result in different orbital hybridization, different orbital compositions, and different band edge energy-positions along different directions in momentum space. We constructed lateral MSJs of metal NbSe$_2$ and semiconductor WSe$_2$ connected along the zigzag and armchair directions to demonstrate the anisotropic band edge energy, orbital character, and SBH. We further show that the anisotropy in momentum space exist also in for other group VIB TMDs. Our finding provides new guidance for the experiments: namely, the direction of electrode wiring is very important for tuning the SBH, or optimizing the electron transport direction may be important. Also, our current work extends the research field of 2D material anisotropy to 2D materials with high symmetry, thus greatly expands the candidate materials for anisotropic studies.

## ACKNOWLEDGMENTS

This work was supported by the National Natural Science Foundation of China (Grants Nos. 12274111 and 12104124), the Central Guidance on Local Science and Technology Development Fund Project of Hebei Province (No. 236Z0601G), the Natural Science Foundation of Hebei Province of China (Nos. A2023201029, B2024201089), the Excellent Youth Research Innovation Team of Hebei University (No. QNTD202412), the Advanced Talents Incubation Program of the Hebei University (Grants Nos. 521000981390, 521000981394, 521000981395, 521000981423, and 521100221055), the Scientific Research and Innovation Team of Hebei University (No. IT2023B03), and the high-performance computing center of Hebei University.

## REFERENCES

[1] X. Chen, Z. Zhou, B. Deng, Z. Wu, F. Xia, Y. Cao, L. Zhang, W. Huang, N. Wang and L. Wang, Electrically tunable physical properties of two-dimensional materials, Nano Today 27, 99 (2019).
[2] Y. Guo, K. Xu, C. Wu, J. Zhao and Y. Xie, Surface chemical-modification for engineering the intrinsic physical properties of inorganic two-dimensional nanomaterials, Chem. Soc. Rev. 44, 637 (2015).
[3] G. Zhang and Y.-W. Zhang, Thermal properties of two-dimensional materials, Chinese Phys. B 26, 034401 (2017).
[4] Q. H. Wang, K. Kalantar-Zadeh, A. Kis, J. N. Coleman and M. S. Strano, Electronics and optoelectronics of two-dimensional transition metal dichalcogenides, Nat. Nanotechnol. 7, 699 (2012).
[5] K. F. Mak, C. Lee, J. Hone, J. Shan and T. F. Heinz, Atomically Thin MoS$_2$: A New Direct-Gap Semiconductor, Phys. Rev. Lett. 105, 136805 (2010).
[6] Z. Lin, C. Wang and Y. Chai, Emerging Group-VI Elemental 2D Materials: Preparations, Properties, and Device Applications, Small 16, 41 (2020).
[7] J. Zhou, J. Lin, X. Huang, Y. Zhou, Y. Chen, J. Xia, H. Wang, Y. Xie, H. Yu, J. Lei, et al., A library of atomically thin

metal chalcogenides, Nature 556, 355 (2018).

[8] H. Ning, Z. Yu, T. Li, H. Shen, G. Long, Y. Shi and X. Wang, From lab to fab: path forward for 2D material electronics, Sci. China Inf. Sci. 66, 160411 (2023).

[9] Q. Wang, J. Tang, X. Li, J. Tian, J. Liang, N. Li, D. Ji, L. Xian, Y. Guo, L. Li, et al., Layer-by-layer epitaxy of multi-layer MoS2 wafers, Natl. Sci. Rev. 9, 6 (2022).

[10] W. Li, X. Gong, Z. Yu, L. Ma, W. Sun, S. Gao, Ç. Köroğlu, W. Wang, L. Liu, T. Li, et al., Approaching the quantum limit in two-dimensional semiconductor contacts, Nature 613, 274 (2023).

[11] P.-C. Shen, C. Su, Y. Lin, A.-S. Chou, C.-C. Cheng, J.-H. Park, M.-H. Chiu, A.-Y. Lu, H.-L. Tang, M. M. Tavakoli, et al., Ultralow contact resistance between semimetal and monolayer semiconductors, Nature 593, 211 (2021).

[12] E. Liu, Y. Fu, Y. Wang, Y. Feng, H. Liu, X. Wan, W. Zhou, B. Wang, L. Shao, C.-H. Ho, et al., Integrated digital inverters based on two-dimensional anisotropic $ReS_2$ field-effect transistors, Nat. Commun. 6, 6991 (2015).

[13] J. Qiao, X. Kong, Z.-X. Hu, F. Yang and W. Ji, High-mobility transport anisotropy and linear dichroism in few-layer black phosphorus, Nat. Commun. 5, 4475 (2014).

[14] P. Li, J. Zhang, C. Zhu, W. Shen, C. Hu, W. Fu, L. Yan, L. Zhou, L. Zheng, H. Lei, et al., Penta-PdPSe: A New 2D Pentagonal Material with Highly In-Plane Optical, Electronic, and Optoelectronic Anisotropy, Adv. Mater. 33, 2102541 (2021).

[15] S. Huang, C. Wang, Y. Xie, B. Yu and H. Yan, Optical properties and polaritons of low symmetry 2D materials, PI 2, R03 (2023).

[16] Z. Zhou, Y. Cui, P.-H. Tan, X. Liu and Z. Wei, Optical and electrical properties of two-dimensional anisotropic materials, J. Semicond. 40, 061001 (2019).

[17] N. Zibouche, S. M. Gunasekera, D. Wolverson and M. Mucha-Kruczynski, Using in-plane anisotropy to engineer Janus monolayers of rhenium dichalcogenides, Phys. Rev. Mater. 6, 054002 (2022).

[18] S. Tang, C. Zhang, D. Wong, Z. Pedramrazi, H.-Z. Tsai, C. Jia, B. Moritz, M. Claassen, H. Ryu, S. Kahn, et al., Quantum spin Hall state in monolayer 1T'-WTe2, Nat. Phys. 13, 683 (2017).

[19] S. Tongay, H. Sahin, C. Ko, A. Luce, W. Fan, K. Liu, J. Zhou, Y.-S. Huang, C.-H. Ho, J. Yan, et al., Monolayer behaviour in bulk $ReS_2$ due to electronic and vibrational decoupling, Nat. Commun. 5, 3252 (2014).

[20] See Supplemental Material at [URL will be inserted by publisher] for the three notes and eight additional figures, which includes Refs.[21, 22].

[21] S. Fang, R. Kuate Defo, S. N. Shirodkar, S. Lieu, G. A. Tritsaris and E. Kaxiras, Ab initio tight-binding Hamiltonian for transition metal dichalcogenides, Phys. Rev. B 92, 205108 (2015).

[22] J. Kang, S. Tongay, J. Zhou, J. Li and J. Wu, Band offsets and heterostructures of two-dimensional semiconductors, Appl. Phys. Lett. 102, 012111 (2013).

[23] R. O. Jones, Density functional theory: Its origins, rise to prominence, and future, Rev. Mod. Phys. 87, 897 (2015).

[24] M. Fuchs and M. Scheffler, Ab initio pseudopotentials for electronic structure calculations of poly-atomic systems using density-functional theory, Comput. Phys. Commun. 119, 67 (1999).

[25] G. Kresse and J. Furthmüller, Efficiency of ab-initio total energy calculations for metals and semiconductors using a plane-wave basis set, Comp. Mater. Sci. 6, 15 (1996).

[26] J. Hafner, Ab-initiosimulations of materials using VASP: Density-functional theory and beyond, J. Comput. Chem. 29, 2044 (2008).

[27] G. Kresse and D. Joubert, From ultrasoft pseudopotentials to the projector augmented-wave method, Phys. Rev. B 59, 1758 (1999).

[28] P. E. Blöchl, Projector augmented-wave method, Phys. Rev. B 50, 17953 (1994).

[29] J. P. Perdew, K. Burke and M. Ernzerhof, Generalized Gradient Approximation Made Simple, Phys. Rev. Lett. 77, 3865 (1996).

[30] D. J. Chadi, Special points for Brillouin-zone integrations, Phys. Rev. B 16, 1746 (1977).

[31] S. Smidstrup, T. Markussen, P. Vancraeyveld, J. Wellendorff, J. Schneider, T. Gunst, B. Verstichel, D. Stradi, P. A. Khomyakov, U. G. Vej-Hansen, et al., QuantumATK: an integrated platform of electronic and atomic-scale modelling tools, J. Phys.-Condens. Mat. 32, 015901 (2019).

[32] V. Wang, N. Xu, J.-C. Liu, G. Tang and W.-T. Geng, VASPKIT: A user-friendly interface facilitating high-throughput computing and analysis using VASP code, Comput. Phys. Commun. 267, 108033 (2021).

[33] K. Momma and F. Izumi, VESTA 3 for three-dimensional visualization of crystal, volumetric and morphology data, J. Appl. Crystallogr. 44, 1272 (2011).

[34] J. E. Padilha, H. Peelaers, A. Janotti and C. G. Van de Walle, Nature and evolution of the band-edge states in $MoS_2$: From monolayer to bulk, Phys. Rev. B 90, 205420 (2014).

[35] V. T. Vu, T. T. H. Vu, T. L. Phan, W. T. Kang, Y. R. Kim, M. D. Tran, H. T. T. Nguyen, Y. H. Lee and W. J. Yu, One-Step Synthesis of $NbSe_2$/Nb-Doped-$WSe_2$ Metal/Doped-Semiconductor van der Waals Heterostructures for Doping Controlled Ohmic Contact, ACS Nano 15, 13031 (2021).

[36] J. Guan, H.-J. Chuang, Z. Zhou and D. Tománek, Optimizing Charge Injection across Transition Metal Dichalcogenide Heterojunctions: Theory and Experiment, ACS Nano 11, 3904 (2017).

[37] Y. An, Y. Hou, K. Wang, S. Gong, C. Ma, C. Zhao, T. Wang, Z. Jiao, H. Wang and R. Wu, Multifunctional Lateral Transition-Metal Disulfides Heterojunctions, Adv. Funct. Mater. 30, 2002939 (2020).

[38] C. Strahberger, J. Smoliner, R. Heer and G. Strasser, Enhanced $k_\parallel$ filtering effects in ballistic electron emission experiments, Phys. Rev. B 63, 205306 (2001).

[39] H. Park, M. Lee, X. Wang, N. Ali, K. Watanabe, T. Taniguchi, E. Hwang and W. J. Yoo, Anisotropic charge transport at the metallic edge contact of $ReS_2$ field effect transistors, Commun. Mater. 5, 87 (2024).

[40] Y. Sun, R. Zhang, J. Tan, S. Zeng, S. Li, Q. Wei, Z.-Y. Zhang, S. Zhao, X. Zou, B. Liu, et al., Tunable in-plane conductance anisotropy in 2D semiconductive $AgCrP2S6$ by ion-electron co-modulations, Sci. Adv. 11, eadr3105 (2025).

[41] M. Aras, Ç. Kılıç and S. Ciraci, Planar heterostructures of single-layer transition metal dichalcogenides: Composite structures, Schottky junctions, tunneling barriers, and half metals, Phys. Rev. B 95, 075434 (2017).